\documentstyle[preprint,prc,aps]{revtex}

\begin{document}
\tighten
\draft
\preprint{
\vbox{
\hbox{TPR--94--02}
\hbox{ADP-94-1/T144}
}}
\title{
Shadowing, Binding and Off-Shell Effects in Nuclear Deep Inelastic Scattering
\footnote{Work supported in part by BMFT grant 06\,OR\,735 and by
Alexander von Humboldt Foundation}
}
\author{S. A. Kulagin\footnote{On leave of absence from
                               Institute for Nuclear Research
                               of the Russian Academy of Sciences,
                               Profsoyusnaya st. 7a, 117312 Moscow, Russia.
                               Alexander von Humboldt fellow.}}
\address{Institut f\"{u}r Theoretische Physik,
         Universit\"{a}t Regensburg,
         D--93040 Regensburg, Germany.}

\author{G. Piller\footnote{On leave from Institute of Theoretical Physics,
                            University of Regensburg, D--93040 Regensburg,
                            Germany}}
\address{Department of Physics and Mathematical Physics,
         University of Adelaide,
         South Australia, 5005.}

\author{W. Weise}
\address{Institut f\"{u}r Theoretische Physik,
         Universit\"{a}t Regensburg,
         D--93040 Regensburg, Germany.}
\maketitle
\newpage
\begin{abstract}
We present a unified description of nuclear deep inelastic scattering (DIS)
over the whole region $0<x<1$ of the Bjorken variable.  Our approach is
based on a relativistically covariant formalism which uses analytical
properties of quark correlators.  In the laboratory frame it naturally
incorporates two mechanisms of DIS: (I) scattering from quarks and
antiquarks in the target and (II) production of quark-antiquark pairs
followed by interactions with the target. We first  calculate structure
functions of the free nucleon and develop a model for the quark spectral
functions. We show that mechanism (II) is responsible for the sea quark
content of the nucleon while mechanism (I) governs the valence part of the
nucleon structure functions. We find that the coherent interaction of $\bar
qq$ pairs with nucleons in the nucleus leads to shadowing at small $x$ and
discuss this effect in detail. In the large $x$ region DIS takes place
mainly on a single nucleon.  There we focus on the derivation of the
convolution model. We point out that the off-shell properties of the bound
nucleon structure function give rise to sizable nuclear effects.

\end{abstract}

\newpage

\section{Introduction}  \label{intro}

Deep-inelastic lepton scattering (DIS) on nuclei is a powerful tool to
investigate the quark-gluon structure of nucleons in a nuclear environment.
The cross-section of this process is studied as a function of the Bjorken
variable $x=Q^2/2M\nu$ and the squared four-momentum transfer $Q^2=-q^2$,
where $M$ is the nucleon mass and $\nu$ is the photon energy in the
laboratory frame. Accurate experimental data
\cite{NMColl92b,NMColl91,E66592,EMColl88,BCDMS87,SLAC14,Arnold84} are now
available for a number of nuclear targets over a wide kinematical range,
$5\cdot 10^{-5}<x<0.8$ and $0.03\:\text{GeV}^2<Q^2<200\:\text{GeV}^2$.  The
data show nontrivial nuclear effects over the whole range of Bjorken $x$.
At $x<0.1$ one observes  shadowing, i.e. a systematic reduction of the
nuclear structure function $F_2^A$ with respect to $A$ times the free
nucleon structure function $F_2^N$. A small enhancement of the ratio
$R=F_2^A/AF_2^N$ is seen  at $x\approx 0.2$ and a pronounced dip occurs in
that ratio at $x\sim 0.5$.  Finally for $x>0.7$ a large enhancement of $R$
is observed.

Numerous models have been proposed to explain these effects (for recent
reviews see e.g.  refs.\cite{BicTho89,CovPre90,Arneod92}).  So far, most
theoretical models for nuclear DIS give separate descriptions of the
regions of small $x<0.1$ and large $x>0.2$. The physical reason for such a
division  becomes apparent in the space-time analysis of the DIS process.
In the laboratory frame the interaction of the virtual photon with the
target can proceed in two possible ways (see Fig.~\ref{fig1}):
\begin{enumerate}
\item[(I)] the photon is absorbed by a quark or antiquark in the target which
picks up the large energy and momentum transfer;
\item[(II)] the photon converts
into a quark-antiquark pair  which subsequently interacts with the target.
\end{enumerate}
An analysis of the contributions  (I) and (II) reveals that the second
mechanism dominates at small $x\ll 0.1$ (see e.g. the discussion in
\cite{PilWei90,BiScSt88,Shaw89,DuBrHo92}).  At $0.1 < x < 0.3$ both
processes (I) and (II) are important, whereas the region $x > 0.3$ is
governed by mechanism (I).  In order to understand the implications of this
for nuclear targets let us consider characteristic space-time scales.
Mechanism (I) has a characteristic scale which  is determined by the size
of the nucleon and does not depend on $x$. For  mechanism (II) the
propagation length $\lambda$ of the $q\bar q$ (or hadronic) fluctuations of
the photon in the laboratory frame is $\lambda\sim (Mx)^{-1}$.  For $x<0.1$
this propagation length becomes larger than the average nucleon-nucleon
distance in nuclei. As a consequence deep-inelastic scattering from nuclear
targets at small $x$ involves the coherent interaction of the pair with
several nucleons in the nucleus. This leads to nuclear shadowing. Models of
the shadowing effect in the laboratory frame
\cite{PilWei90,BiScSt88,Shaw89,FraStr89,BrodLu90,NikZak91,BadKwi92b,MelTho93}
usually include mechanism (II) only.

At large $x>0.2$ effects resulting from coherent multiple scattering in the
nucleus are not important since the space-time scales of both mechanisms
(I) and (II) are of the order of the nucleon size.  In this region of $x$
the virtual photon interacts incoherently with bound nucleons.  Model
descriptions of nuclear structure functions in the large $x>0.2$ region
\cite{AkKuVa85,BiLeSh89,DunTho86,Kulagi89,JunMil88,CioLiu89,DieMil91,BGNiPZ93}
(for a review see \cite{BicTho89,Jaffe85}) usually start out from the
impulse approximation in which both mechanisms (I) and (II) are taken into
account by using a phenomenological nucleon structure function.  Results of
such calculations show that nuclear binding and Fermi motion are
responsible for the observed ``old'' EMC-effect at large $x$.

The purpose of the present paper is to develop an approach to DIS which is
based on a unified description of processes (I) and (II).  In
Section~\ref{frame} we develop a relativistically covariant formalism which
incorporates the standard parton model but also permits us to include
non-perturbative features which turn out to be important at small $x$.  Our
starting point is a general representation of structure functions in terms
of dispersion integrals over quark spectral densities (Section~\ref{DI}).
In Section~\ref{model} we develop a model for the quark spectral densities
which separately reproduces the valence and sea quark parts of the free
nucleon structure functions. In Section~\ref{kern} we discuss nuclear
structure functions.  We find that shadowing at small $x$ has a scaling
contribution from mechanism (II) (independent of $Q^2$) which turns out to
be insufficient, however, to reproduce the empirical $A$-dependence. At
this point our results differ from those of ref.\cite{BrodLu90}. We
conclude that the propagation of strongly correlated $q\bar q$ pairs
through the nucleus, partly in the form of vector mesons, is important to
reproduce the observed shadowing effect (Section~\ref{smallx}).  For
$x>0.2$, where nuclear binding and Fermi motion are relevant, we discuss
the limitations of the standard convolution model (Section~\ref{largex}).
We point out that there is no reason to ignore, as is usually done, the
dependence of the structure functions on the invariant mass of the nucleon,
$p^2$.

\section{Framework} \label{frame}

According to the optical theorem inclusive inelastic scattering of an
electron/muon on a nucleon or nucleus can be described in terms of
the forward scattering of a virtual photon.
The amplitude for forward Compton scattering is
\begin{eqnarray}
\label{Compton}
  T_{\mu\nu}(P,q)=-i\int d^{4}\xi\: e^{iq\cdot\xi}
        \langle P|T\left(j_{\mu}(\xi)j_{\nu}(0)\right)|P \rangle ,
\end{eqnarray}
where $q$ and $P$ are the photon and target momenta, respectively. The
electromagnetic current is denoted by $j_{\mu}$. In what follows we shall
discuss the scattering from an unpolarized target and assume that the average
is taken over target polarization in eq.(\ref{Compton}). In this case there
are only two independent terms in the Compton amplitude,
\begin{equation}
\label{T1-2}
    T_{\mu\nu}(P,q)=T_1(x,Q^2)
        \left(\frac{q_{\mu}q_{\nu}}{q^2}-g_{\mu\nu}\right)
            +\frac{T_2(x,Q^2)}{P\!\!\cdot\!q}L_{\mu}L_{\nu},
\end{equation}
where  $Q^2=-q^2$, $L_{\mu}=P_{\mu}\!-\!q_{\mu}P\!\!\cdot\!q/q^2$,
and $x=Q^2/2P\!\!\cdot\!q$ is the Bjorken scaling variable (we use the
normalization $\langle p|p'\rangle=(2\pi)^3\, 2p_0\delta^{(3)}({\bf p}-{\bf
p'})$ both for fermions and bosons, so that the amplitudes $T_1$ and $T_2$
are dimensionless).  The structure functions $F_1$ and $F_2$ are given by
the imaginary parts of the scalar amplitudes in (\ref{T1-2}):
\begin{equation}    \label{F1-2}
    F_{1,2}(x,Q^2)=-\frac{1}{2\pi}\,{\rm Im} T_{1,2}(x,Q^2)  .
\end{equation}

It is commonly assumed that in the region of high momentum transfer,
$Q^2\gg M^2$, the main contribution to the Compton amplitude comes from the
diagram, Fig.~\ref{fig2}, in which the virtual photon couples to the quark
current.  Let us examine this contribution in detail.  To leading order in
$Q^2$ and in the axial gauge the Compton amplitude reads
\begin{eqnarray} \label{IA}
T_{\mu\nu}(P,q)=-i\int \frac{d^4k}{(2\pi)^4} \sum_a e_a^2
               \: {\rm Tr}\left[\Delta_a(k,P)\left(\gamma_{\mu}
\frac{1}{\overlay{\slash}k + \overlay{\slash}q+i\epsilon}\gamma_{\nu} +
\gamma_{\nu}\frac{1}{\overlay{\slash}k-\overlay{\slash}q+i\epsilon}
         \gamma_{\mu}\right)\right],
\end{eqnarray}
where  the sum is taken over flavor and color degrees of freedom of the
interacting quark which carries  electric charge $e_a$.  For simplicity we
have dropped the quark mass in eq.(\ref{IA}).  Here $\Delta_a(k,P)$ is the
Fourier transform of the correlator of  the quark fields in  the target,
\begin{eqnarray}
\label{Delta}
    \Delta_a(k,P)= -i\int d^{4}\xi\: e^{ik\cdot\xi} \langle
    P|T\left(\psi_a(\xi)\bar{\psi}_a(0)\right)|P \rangle  .
\end{eqnarray}
The structure functions can easily be found from eq.(\ref{IA}) by applying
appropriate projection operators.  Neglecting terms of order  $1/Q^2$ we
find for the structure function $F_2$,
\begin{eqnarray} \label{F2IA}
    F_2(x) =x\sum_a e_a^2\,\left(q_a(x)+\bar q_a(x)\right) ,
\end{eqnarray}
where \vspace*{-\baselineskip}\vspace*{-\abovedisplayskip}
\begin{mathletters}%
\label{qqbar}
\begin{eqnarray}
q_a(x)      &=& f_a(x),\\
\bar q_a(x) &=& -f_a(-x),
\end{eqnarray}
\end{mathletters}%
and\vspace*{-\baselineskip}\vspace*{-\abovedisplayskip}
\begin{eqnarray}
\label{f}
f_a(x) &=& -i\int \frac{d^4k}{(2\pi)^4}
\frac{{\rm Tr}\left(\overlay{\slash}q\Delta_a(k,P)\right)}{2P\!\cdot\! q}
                \delta\left(x - \frac{k\!\cdot\! q}{P\!\cdot\! q}\right)  .
\end{eqnarray}
The two terms in eq.(\ref{F2IA}) correspond to  the  direct and crossed
terms of the Compton amplitude. It follows from eq.(\ref{IA}) that the
structure function $F_1$ is not independent; it is given by the
Callan-Gross relation, $F_2(x)=2x\,F_1(x)$.

\subsection{Parton model}

Equation (\ref{f}) gives a Lorentz covariant representation of the quark
distribution function which can be used in any reference frame.  Here we
first demonstrate that eqs.(\ref{F2IA},\ref{f}) recover the familiar result
of the parton model. Let us choose a reference frame in which the target
moves with a large momentum $|{\bf P}|\rightarrow\infty$. In this frame the
function $q_a(x)$ can be identified with the momentum distribution of
quarks with flavor $a$ in the target, and $\bar q_a(x)$ is the
corresponding antiquark distribution.  In order to see this we introduce a
coordinate system such that the  momentum transfer is $q=(0,{\bf
0}_{\perp},Q)$ (with $Q=\sqrt{Q^2}$). The hadron moves with momentum
$P_3=-Q/2x$ in the direction opposite to the three-momentum transfer ${\bf
q}$.  Then eq.(\ref{f}) becomes
\begin{equation} \label{fIMF}
    f(x)=-i\int\!\frac{d^3k}{(2\pi)^3}
        \delta\left(x\!-\!\frac{k_3}{P_3}\right)
        \int\frac{dk_0}{2\pi}
    \frac{{\rm Tr}(\gamma_3\Delta(k,P))}{2P_3} ,
\end{equation}
where we have suppressed quark flavor and color indices for simplicity.
It can be shown that $\gamma_3/P_3=\gamma_0/P_0$ in eq.(\ref{fIMF}), up to
terms of order $M^2x^2/Q^2$.  We perform the  $k_0$-integration by closing
the integration  contour in the upper half-plane and obtain:
\begin{equation} \label{N1}
      -i\int\frac{dk_0}{2\pi}\frac{{\rm Tr}(\gamma_0\Delta(k,P))}{2P_0}=
            \int d^3r\,d^3r'e^{-i{\bf k\cdot}({\bf r}-{\bf r'})}
        \langle
    \psi^{\dagger}(0,{\bf r'})\psi(0,{\bf r})
        \rangle .
\end{equation}
Here we have used translational  invariance and introduced an additional
space integration.  The brackets in eq.(\ref{N1}) denote
$\langle\cdots\rangle=\langle P|\cdots|P \rangle/\langle P|P\rangle$.
The r.h.s. of eq.(\ref{N1}) is in fact the momentum space density of
quarks in the target. In order  to clarify this connection further we expand
the quark fields in terms of plane wave spinors,
\begin{eqnarray} \label{momspace}
        \psi(0,{\bf r})=\sum_{\sigma}\!\int\!\frac{d^3k}{(2\pi)^3}
  \frac{1}{\sqrt{2|{\bf k}|}}
  \left(
a(k,\sigma)\mbox{\large\it u}(k,\sigma)e^{i{\bf k\cdot r}}\!+
\!b^{\dagger}(k,\sigma)\mbox{\large\it v}(k,\sigma)e^{-i{\bf k\cdot r}}
  \right) ,
\end{eqnarray}
where {\large\it u}$(k,\sigma)$ is the Dirac spinor of a quark with
momentum $k$ and polarization $\sigma$, and {\large\it v}$(k,\sigma)$ is
the corresponding antiquark spinor.  The operators $a(k,\sigma)$ and
$b(k,\sigma)$ acting on any physical state probe the momentum distribution
of quarks and antiquarks in that state:
\begin{mathletters}%
\begin{eqnarray}
N(k)      &=& \sum_{\sigma}\langle a^{\dagger}(k,\sigma) a(k,\sigma)\rangle,
\\
\bar N(k) &=& \sum_{\sigma}\langle b^{\dagger}(k,\sigma) b(k,\sigma)\rangle.
\end{eqnarray}
\end{mathletters}%
Using the commutation relations between $a$, $a^{\dagger}$ and $b$,
$b^{\dagger}$ and the orthogonality properties of spinors {\large\it
u}$(k,\sigma)$ and {\large\it v}$(k,\sigma)$, we find from
eqs.(\ref{fIMF},\ref{N1}),
\begin{eqnarray} \label{fPM}
f(x)= q(x)-\bar q(-x),
\end{eqnarray}
where $q(x)$ and $\bar q(x)$ are the quark and antiquark distributions as
functions of the target momentum fraction,
\begin{mathletters}%
\label{PM:qqbar}
\begin{eqnarray}
    q(x)&=&\int\!\frac{d^3k}{(2\pi)^3}
    \delta\!\left(x-\frac{k_3}{P_3}\right)\, N({\bf k}) , \\
    \bar q(x)&=&\int\!\frac{d^3k}{(2\pi)^3}
    \delta\!\left(x-\frac{k_3}{P_3}\right)\, \bar N({\bf k}) .
\end{eqnarray}
\end{mathletters}%
The functions (\ref{PM:qqbar}) have the usual simple
interpretation: they represent the probability distributions of quarks
($q(x)$) or antiquarks ($\bar q(x)$) which carry a fraction $x$ of the
target longitudinal momentum.  It is known from the parton model
analysis\cite{IoKhLi84} that the probability to find a parton moving
backward ($x<0$) vanishes as $|{\bf P}|\rightarrow\infty$. Also, momentum
conservation does not permit partons with $x>1$. Therefore the  functions
$q(x)$ and $\bar q(x)$ vanish outside the physical interval $0\le x\le 1$.

\subsection{Structure functions as dispersion integrals}    \label{DI}

Our primary task is to study deep-inelastic scattering off nuclei. In this
case the simplicity of the parton model is lost, because no reliable
approach exists to deal with nuclear systems in the infinite momentum
frame.  For this purpose the preferable frame of reference is the
laboratory system. In the present paper we describe the distribution
function $f(x)$ using the analytical properties of the quark correlator
(\ref{Delta}).  This method preserves relativistic covariance and can
therefore be used in any frame.

Following ref.\cite{LaPoSh71} we assume that the quark correlator $\Delta$
is an analytic function of the variables $s=(P-k)^2$, $u=(P+k)^2$ and
$k^2$.  For real $s$ and $u$ the quark correlator has a right-hand  cut in
the variable $s$, a left-hand cut in the variable $u$ and singularities for
$k^2>0$.  In order to make use of these analytical properties of $\Delta$
in the loop integral (\ref{f}), it is convenient to parameterize the loop
momentum $k$ in terms of the external momenta $P$ and $q$ (introducing the
Sudakov variables):
\begin{equation}    \label{sudakov}
k=\alpha P + \beta q' + k_{\perp}  ,
\end{equation}
where $q'=x\,P+q$, and $k_{\perp}$ is a two-dimensional vector
(with $k_{\perp}^2<0$) perpendicular to both $P$ and $q$.

The integration with respect to $\beta$ can be done using the analytic
properties, just mentioned, of the quark correlator. As a result one finds
that the distribution function $f(x)$ vanishes outside the physical interval
$|x|\le 1$ as it should. For $0\le x\le 1$ the distribution function $f(x)$
(or $q(x)$) is given by a dispersion integral in the variable $s$ along the
right-hand cut.  For $-1\le x\le 0$ the $u$-channel cut is relevant.  In
order to represent the structure functions in this way we introduce functions
$\rho_{R,L}$ for the imaginary parts taken along the right-hand cut ($R$) and
left-hand cut ($L$) as follows:
\begin{mathletters}%
\label{rho}
\begin{eqnarray}
\rho_R(s,k^2,\alpha)    &=&
\frac{{\rm Im}_R\,{\rm Tr}\left(\overlay{\slash}q\Delta(k,P)\right)}
     {2\pi\,P\!\cdot\! q}, \\
\rho_L(u,k^2,\alpha)    &=&
\frac{{\rm Im}_L\,{\rm Tr}\left(\overlay{\slash}q\Delta(k,P)\right)}
     {2\pi\,P\!\cdot\! q} .
\end{eqnarray}
\end{mathletters}%
In terms of these functions the quark distribution $q(x)$ and the antiquark
distribution $\bar q(x)$ are
\begin{mathletters}%
\label{qqbar:DI}
\begin{eqnarray}
q(x)        &=& \frac{1}{1-x}\int \frac{ds\, d^2k_{\perp}}{2(2\pi)^3}\:
          \rho_R(s,k^2,x)  ,\\
\bar q(x)   &=& \frac{-1}{1-x}\int \frac{du\, d^2k_{\perp}}{2(2\pi)^3}\:
          \rho_L(u,k^2,-x),
\end{eqnarray}
\end{mathletters}%
where we have suppressed the flavor indices again. The squared quark
four-momentum $k^2$ is
\begin{equation} \label{ksqr}
k^2=x\left(\frac{s}{x-1}+M^2\right)+\frac{k_{\perp}^2}{1-x}
\end{equation}
in eq.(\ref{qqbar:DI}a), and an analogous expression holds with $s$
replaced by $u$ for the antiquark distribution $\bar q(x)$ in
eq.(\ref{qqbar:DI}b).

The spectral densities (\ref{rho}) can be written in terms of spectral
sums over a complete set of intermediate states inserted between the two quark
field operators in eq.(\ref{Delta}).  In order to write the spectral
representation in a more explicit form we introduce matrix elements of the
quark field operator taken between the nucleon  and some intermediate state,
$\psi_n({\bf K})=\langle {\bf K}, n|\psi(0)|P\rangle$
and
$\tilde\psi_m({\bf K})=\langle P|\psi(0)|m,{\bf K} \rangle$.
Intermediate states are labeled by their total momentum ${\bf K}$ and
other quantum numbers denoted by $n$ or $m$. In terms of the  amplitudes
$\psi_n$ and $\tilde\psi_m$ the spectral densities (\ref{rho}) can then be
expressed as follows:
\begin{eqnarray}
\label{rho-R}
\rho_R(s,k^2,\alpha) &=& \frac{1}{2P\!\cdot\!q}
\sum_n \overline{\psi}_n({\bf P\!-\!k})\overlay{\slash}q\psi_n({\bf P\!-\!k})
            \delta(s-M_n^2) ,\\
\label{rho-L}
-\rho_L(u,k^2,-\alpha) &=& \frac{1}{2P\!\cdot\!q}
    \sum_m \overline{\tilde\psi}_m({\bf P\!-\!k})\overlay{\slash}q
            \tilde\psi_m({\bf P\!-\!k})
                \delta(u-M_m^2) .
\end{eqnarray}
Here $M_n^2$ and $M_m^2$ are the invariant masses of the intermediate
states.  Note that eqs.(\ref{rho-R}) and (\ref{rho-L}) correspond to two
different time orderings of the quark operators in (\ref{Delta}).

The basic assumption is now that the spectral densities vanish at large
$k^2$ so that integrals in eqs.(\ref{qqbar:DI}) are convergent and
dominated by the region of $k^2\sim {\bar m}^2$, where ${\bar m}$ is a
characteristic hadronic mass scale.  It follows from eq.(\ref{ksqr}) that
the behavior of the quark distributions at $x\to 1$ is given by the
asymptotics of the spectral densities at $k^2\to -\infty$.  At small $x$
the $k^2$ is finite even for large $s\sim {\bar m}^2/x$. Therefore for
small $x$ the integral in eqs.(\ref{qqbar:DI}) is sensitive to the
high-energy parts of the spectral densities. In the region of intermediate
or large $x>0.2$ the region of finite $s\sim {\bar m}^2$ is of major
importance.

The spectral representation (\ref{rho-R},\ref{rho-L}) offers a convenient
way to separate contributions from mechanisms (I) and (II).  To see this,
consider the amplitude $\psi_n$ in the laboratory frame, with target
momentum $P=(M,{\bf 0})$.  The quark field operator acting on the target
state can either annihilate a quark in the target or create an antiquark.
In the former case the amplitude $\psi_n\propto \langle -{\bf k},
n|a(k)|P\rangle$ describes the absorption of a virtual photon by quarks
with momentum ${\bf k}$.  This contribution corresponds to the mechanism
(I).  The contribution from the antiquark part of the $\psi$-operator,
$\psi_n\propto \langle -{\bf k}, n|b^{\dagger}(-k)|P\rangle$, corresponds
to the mechanism (II). This part of the amplitude $\psi_n$ describes the
"external" antiquarks from $q\bar q$ fluctuations of the virtual photon and
their interaction with the target.  The amplitude $\tilde \psi_m$ also has
two parts, one associated with contributions from antiquarks bound to the
target ($\tilde\psi_m^{\dagger}\propto\langle {\bf -k}, m|b(k)|P\rangle$)
and the other one from "external" quarks coming from the photon wave
function,
($\tilde\psi_m^{\dagger}\propto\langle {\bf -k}, m|a^{\dagger}(-k)|P\rangle$).

\section{Quark Spectral Densities and Nucleon Structure Functions}
\label{model}

In this section we construct nucleon structure functions using a model for
the quark spectral densities $\rho_L$ and $\rho_R$.  We further elaborate
the concept, appropriate in the laboratory frame, that the full quark
spectral function can be divided into two parts corresponding to mechanisms
(I) and (II), as illustrated in Fig.~\ref{fig1}.

The part of the spectral density that describes mechanism (I) is
proportional to the probability to find a quark with four-momentum $k$ in
the nucleon. The characteristic momenta of quarks bound in the nucleon are
of the order of the nucleon mass, $k\sim M$. Therefore the main
contributions to the spectral densities from mechanism (I) come from the
region $|k^2|\sim M^2$ and $s\sim M^2$.  This kinematical region determines
the behavior of structure functions at $x>0.2$.

On the other hand, contributions to the quark spectral densities from
mechanism (II) rise with $s$ (or $u$). In fact, the matrix element
$\langle{\bf k}, n|b^{\dagger}(k)|P\rangle$ describes scattering of an
antiquark with momentum ${\bf k}$ from the nucleon, with transition of the
system to the final state $|n,{\bf k}\rangle$.  The possible antiquark
momenta ${\bf k}$ are determined by the wave function of the photon and can
be as large as the photon momentum ${\bf q}$. Therefore, the invariant mass
of the antiquark-nucleon system is large, $s\gg M^2$.  Due to  unitarity,
the sum over all states $n$ in eq.(\ref{rho-R}) will be proportional to the
antiquark-nucleon forward elastic scattering amplitude. It is known from
Regge theory that imaginary parts of amplitudes rise with energy as
$s^{\alpha_P}$ where $\alpha_P$ is the intercept of the Pomeron.  Therefore
contributions to the spectral density from  mechanism (II) grow with energy
and dominate at large $s$.  This mechanism determines the small $x$ part of
structure functions.

\subsection{Model for quark spectral densities\label{model:spectra}}

With this discussion in mind we now develop the following simple model for
the quark spectral densities. We introduce a parameter $s_0$ which
separates the full spectrum into a low-energy $(s<s_0)$ part and a
high-energy $(s>s_0)$ part. We assume furthermore that the low-energy part
of the spectrum is dominated by  mechanism (I), while the high-energy part
is given by  mechanism (II) (for illustration see Fig.~\ref{fig3}):
\begin{equation}
\label{spectrum}
\rho(s,k^2)=\rho^{{\rm (I)}}(s,k^2)\theta(s_0-s)+
\rho^{{\rm (II)}}(s,k^2)\theta(s-s_0) .
\end{equation}
For the low-energy part of the spectrum ($s<s_0$) we make the following
ansatz:
\begin{eqnarray} \label{spectrum:I}
\rho_R^{{\rm (I)}}(s,k^2) = \Phi(k^2)\delta(s-\bar s),
\end{eqnarray}
together with $\rho_L=0$ in this region, which implies that we neglect
contributions to spectral densities coming from antiquarks in the nucleon.
This choice can be motivated within a constituent quark picture of the
nucleon. In this case $\sqrt{\bar s}\approx {2\over 3}M$ is an average
mass of the residual two-quark intermediate state, and $\Phi(k^2)$ is the
squared momentum space wave function of the constituent quark.  Here we
will not confine ourselves to some particular model but rather choose the
spectral density at small values of $s$ in such a  way that we reproduce
the measured valence quark distribution.  The contribution from
eq.(\ref{spectrum:I}) to the deep-inelastic structure function is,
\begin{equation} \label{qI}
q^{{\rm (I)}}(x)=\frac{1}{16\pi^2}
\int_{-\infty}^{k^2_{{\rm max}}(\bar s,x)} \!\!dk^2\: \Phi(k^2) ,
\end{equation}
where $k^2_{{\rm max}}(s,x)$ is the maximum value of the squared
quark four-momentum  for given values of $s$ and $x$,
\begin{equation} \label{k2max}
k^2_{{\rm max}}(s,x)=x\left(\frac{s}{x - 1}+M^2\right) .
\end{equation}

In order to describe the region of large $s>s_0$ we introduce the
quark-nucleon forward scattering amplitude $T(k,P)$ as follows:
\begin{equation}
\Delta(k,P)=
\frac{1}{\overlay{\slash}k-m_q+i\epsilon}\, T(k,P)\,
\frac{1}{\overlay{\slash}k-m_q+i\epsilon} .
\end{equation}
The quantity relevant for the calculation of the quark spectral densities
(\ref{rho}) is ${\rm Tr}(\gamma_{\mu}\Delta)$. In terms of the amplitude $T$
this trace can be written as follows,
\begin{equation}    \label{Tr-T}
{\rm Tr}(\gamma_{\mu}\Delta(k,P))=(k^2-m_q^2)^{-2}
\left[
4k_{\mu}\!\cdot\!\overline{T}+(m_q^2-k^2)\,{\rm Tr}(\gamma_{\mu}T)
\right],
\end{equation}
where $\overline{T}={1\over 2}{\rm Tr}[(\overlay{\slash}k+m_q)T(k,P)]$
is the amplitude averaged over the quark spin. When examining the Dirac
structure of $T$, we find that only a scalar term and a term proportional
to the $\gamma$-matrix contribute to (\ref{Tr-T}). One can neglect the
scalar term because its contribution to the amplitude $\overline{T}$ is
proportional to the quark mass $m_q$.  The leading contribution to the
amplitude $\overline{T}$, the one which rises with $s$, comes from the
term proportional to $\gamma\!\cdot\!P$. Based on these arguments we write
the amplitude $T$ as
\begin{equation}
\label{T}
T(k,P)=C(s,u,k^2)\,\overlay{\slash}P ,
\end{equation}
where $C$ is a Lorentz invariant function of $s$, $u$ and $k^2$ which is
related to the spin-averaged amplitude $\overline{T}$ as follows:
\begin{equation}    \label{Tbar}
\overline{T}(s,u,k^2)={1\over 2}(u-s)C(s,u,k^2).
\end{equation}
We emphasize here that the amplitude $\overline{T}(s,u,k^2)$ describes both
the $qN$- and the $\bar qN$-scattering channels. In the $s$-channel
$\overline{T}(s,u,k^2)$ coincides with the antiquark-nucleon scattering
amplitude $T_{\bar qN}(s,k^2)$, while in the $u$-channel
$\overline{T}(s,u,k^2)$ gives the quark-nucleon amplitude $T_{qN}(u,k^2)$.

We are now prepared to calculate the contribution from mechanism (II) to
the quark spectral densities (\ref{rho}) and to the quark and antiquark
distributions (\ref{qqbar:DI}).  Using
eqs.(\ref{Tr-T},\ref{T},\ref{Tbar}) we obtain:%
\footnote
{
As compared to the spinless case discussed in \cite{BrodLu90}, the spin
$1/2$ distributions have a generic factor of 2 which reflects the number
of spin degrees of freedom, and the term in  brackets under the integral
replaced the factor  $x$.
}
\begin{mathletters}%
\label{qqbarII}
\begin{eqnarray}
q^{{\rm (II)}}(x) &=& \frac{1}{(2\pi)^3}
    \int_{s_0}^{\infty} ds\int_{-\infty}^{k^2_{{\rm max}}(s,x)}dk^2
        \:\frac{{\rm Im} T_{\bar qN}(s,k^2)}{(k^2-m_q^2)^2}
    \left(-x+\frac{m_q^2-k^2}{s-M^2-k^2}\right) ,\\
\bar q^{{\rm (II)}}(x) &=& \frac{1}{(2\pi)^3}
    \int_{s_0}^{\infty} du\int_{-\infty}^{k^2_{{\rm max}}(u,x)}dk^2
        \:\frac{{\rm Im} T_{qN}(u,k^2)}{(k^2-m_q^2)^2}
    \left(-x+\frac{m_q^2-k^2}{u-M^2-k^2}\right) ,
\end{eqnarray}
\end{mathletters}%
where we have introduced an integration over the squared quark four-momentum
$k^2$ instead of integration over transverse momentum $k_{\perp}$.

In what follows we shall consider the structure function $F_2$ of an
isoscalar nucleon, $F_2^N=\frac{1}{2}(F_2^p + F_2^n)$. Isospin symmetry
implies that $F_2^N$ is proportional to the flavor singlet combination of
quark and antiquark distributions,
\begin{equation}
\label{F2N}
F_2^N(x) = {5\over 18} x \sum_a\left(q_a(x)+\bar q_a(x)\right) .
\end{equation}
(Here we have assumed that the difference between strange and charmed
sea is negligible small). In our model the quark and antiquark distributions
are given by
\begin{equation}
\label{sum-q}
\sum_a\left(q_a(x)+\bar q_a(x)\right) =
q^{{\rm (I)}}(x)+N_fN_c\left(q^{{\rm (II)}}(x)+\bar q^{{\rm (II)}}(x)\right),
\end{equation}
where $q^{{\rm (I)}}$ is given by eq.(\ref{qI}) and represents the sum of
quark distributions of different flavors due to the mechanism (I).  The
quantities $q^{{\rm (II)}}$ and $\bar q^{{\rm (II)}}$ are the quark and
antiquark distributions related to mechanism (II), averaged over flavor
and color. They are given by eqs.(\ref{qqbarII}), where $T_{qN}$
and $T_{\bar qN}$ are quark-nucleon amplitudes averaged over quark spin,
flavor and color. We have $N_f=4$ and $N_c=3$.

The valence quark distribution is measured in neutrino scattering in
terms of the structure function $F_3(x)$. In our model it is given by
\begin{eqnarray}
\label{F3N}
F_3^N(x)=q^{{\rm (I)}}(x) +
N_f N_c\left(q^{{\rm (II)}}(x)-\bar q^{{\rm (II)}}(x)\right) .
\end{eqnarray}
This structure function is normalized to the number of valence
quarks in the nucleon.

\subsection{Quark-nucleon amplitude}

In order to specify $q^{\rm (II)}$ and $\bar q^{\rm (II)}$ we note that
the amplitudes $T_{qN}$ and $T_{\bar qN}$ can be connected to observable
proton-proton and antiproton-proton forward scattering amplitudes. We
recall the well known phenomenological fact that total hadronic
cross-sections at high energies are proportional to the number of
constituent quarks in hadrons\cite{IoKhLi84}.  Hence the forward
proton-proton amplitude can be written in terms of the quark-proton
amplitude in the laboratory frame as:
\begin{equation} \label{Tav}
\frac{T_{pp}(S)}{S}=3\left\langle
\frac{T_{qp}(y S,k^2)}{y S}
\right\rangle .
\end{equation}
Here $S$ is the squared proton-proton center of mass energy, $k$ is the
four-momentum of a constituent quark in the target and $y$ is the fraction
of the target light-cone momentum carried by the constituent quark.  A
similar equation relates the antiquark-proton amplitude $T_{\bar qp}$ to
the antiproton-proton forward scattering amplitude $T_{\bar p p}$.  In the
laboratory frame $y=(k_0 + k_3)/M$, $k_0=M-\sqrt{m_R^2+{\bf k}^2}$, where
$m_R$ is the mass of the spectator system.  The averaging in eq.(\ref{Tav})
is performed over the spectral function of the constituent quarks in the
proton target.  In order to estimate typical values of $y$ and $k^2$ we
assume $m_R\approx\frac23 M$. This gives average values $\bar y\approx0.3$
and $\overline{k^2}=\langle k_0^2-{\bf k}^2\rangle\approx -0.1\:$GeV$^2$.

The constituent quark-nucleon scattering amplitudes in (\ref{Tav}) might in
principle be different from the quark-nucleon amplitudes entering in
eqs.(\ref{qqbarII}).  However, at small values of $x$, the quark-nucleon
center of mass energy $s$ in eq.(\ref{qqbarII}) will be of the order of
$s\sim 1\,$GeV$^2/x$. As a consequence the typical formation time of a
constituent quark will be $\tau_F\sim |k_0|/m_q^2 \sim s/2m_q^2M \sim
1/Mx$. This is comparable to the propagation length $\lambda\sim 1/Mx$ of
the $q\bar q$ pair in the photon wave function.  We can therefore assume
that at small $x$ the quark-nucleon amplitudes which enter in the quark
distribution functions can be approximated by the constituent quark-nucleon
amplitudes determined by hadron-nucleon scattering.

Through eq.(\ref{Tav})  the $s$-dependence of the quark-nucleon and
antiquark-nucleon amplitudes $T_{qN}$ and $T_{\bar qN}$ is fixed. Above
the resonance region the $pp$- and $\bar p p$-amplitudes are well
reproduced by Pomeron exchange with an intercept $\alpha_P=1+\epsilon$,
and by exchange of two Regge trajectories corresponding to the $\rho$- and
the $a_2$-mesons with intercepts $\alpha_{\rho} = \alpha_{a_2} =
\alpha_R\approx 1/2$.  The forward scattering amplitudes can then be
written as \cite{Collin77}:
\begin{mathletters}%
\label{pp}
\begin{eqnarray}
T_{pp}(S) &=& R_P S^{\alpha_P}(i+\tan\frac{\pi\epsilon}{2})+
S^{\alpha_R}\left(iR_{\Delta}-
\frac{R_{\Sigma}+R_{\Delta}\,\cos\pi\alpha_R}{\sin\pi\alpha_R}\right) ,\\
T_{\bar p p}(S) &=& R_P S^{\alpha_P}(i+\tan\frac{\pi\epsilon}{2})+
S^{\alpha_R}\left(iR_{\Sigma}-
\frac{R_{\Delta}+R_{\Sigma}\,\cos\pi\alpha_R}{\sin\pi\alpha_R}\right) ,
\end{eqnarray}
\end{mathletters}%
were $R_P$ is the residue of the Pomeron while $R_{\Sigma}$ and
$R_{\Delta}$ stand for the sum and the difference of residues of the
$\rho$- and $a_2$-trajectories. In the following we shall use the best fit
parameters of ref.\cite{DonLan92}: \[
\begin{array}{ll}
\epsilon = 0.0808, & R_P = 21.70\,\mbox{mb/GeV}^{2\epsilon},  \\
\alpha_R=0.5475,   &
R_{\Delta} = 56.08\,\mbox{mb$\cdot$GeV}^{(1-\alpha_R)},      \\
           & R_{\Sigma} = 98.39\,\mbox{mb$\cdot$GeV}^{(1-\alpha_R)}.
\end{array} \]

The amplitudes $T_{qN}$ and $T_{\bar qN}$ should depend not only on $s$,
but also on the squared quark four-momentum $k^2$. We assume that the
$k^2$-behavior of the Pomeron and  Reggeon parts of the amplitudes is
given as:
\begin{equation} \label{T(k2)}
T_{qN}(s,k^2)=g_P(k^2)T_{qN}^P(s,0) + g_R(k^2)T_{qN}^R(s,0),
\end{equation}
with functions $g_{P,R}(k^2)$ for which we chose the following ansatz:
\begin{equation}    \label{g(k2)}
g_{P,R}(k^2) = (1-k^2/\Lambda_{P,R}^2)^{-n_{P,R}} ,
\end{equation}
with momentum space cutoffs $\Lambda_P$, $\Lambda_R$ and exponents $n_P$
and $n_R$.

\subsection{Nucleon structure functions at large $Q^2$}

In this section we determine $g_P(k^2)$ and $g_R(k^2)$ and other remaining
parameters such that we reproduce the measured structure functions in the
scaling region. This involves the antiquark distributions $\bar q(x)$
(eq.(\ref{qqbarII}) in our model) together with the structure functions
$F^N_2(x)$ and $F^N_3(x)$ (eqs.(\ref{F2N},\ref{F3N})).

The valence quark distribution is mainly given in terms of the momentum
distribution $\Phi(k^2)$ in eq.(\ref{spectrum:I}) which still needs to be
specified. We use the ansatz
\begin{equation}    \label{Phi}
\Phi(k^2) = (1-k^2/\Lambda_V^2)^{-n_V} ,
\end{equation}
with a suitable cutoff parameter $\Lambda_V$ and a characteristic exponent
$n_V$.

{}From eqs.(\ref{qqbar:DI},\ref{ksqr}) it follows that the asymptotic
behavior of the quark spectral functions at $k^2\to-\infty$ determines the
structure functions at $x\to 1$. Indeed, our model gives
\begin{mathletters}%
\label{x-to-1}
\begin{eqnarray}
q^{\rm (I)}(x\to 1)  &\propto& (1-x)^{n_V-1},\\
q^{\rm (II)}(x\to 1) &\propto& (1-x)^{n_{P,R}+1} .
\end{eqnarray}
\end{mathletters}%
The quark counting rule \cite{IoKhLi84,MaMuTa73,BroFar73} for the valence
distribution requires $q^{\rm (I)}(x\to 1)\propto (1-x)^3$ which fixes
$n_V=4$. Furthermore, the sea quark distribution should approach zero at
$x\to 1$ with a high power in $(1-x)$. We find that the Pomeron and Reggeon
exponents $n_P=n_R=4$, which correspond to a $(1-x)^5$ behavior at large
$x$, give a good fit to $\bar q(x)$.

The Regge parameters of the quark-nucleon and antiquark-nucleon
amplitude in eqs.(\ref{qqbarII}) are already fixed at an averaged
squared momentum $|k^2|\sim 0.1\,$GeV$^2$ by employing eq.(\ref{Tav}).
Note that the overall magnitudes of the type (II) quark and antiquark
distributions (\ref{qqbarII}) are set by the $qN$ and $\bar qN$ cross
sections. Using eq.(\ref{Tav}) we find $\sigma_{\bar
qN}\approx\sigma_{qN}\approx \frac13\sigma_{pp}\approx 13\,$mb at high
energy. The remaining scales enter through the cutoffs in $g_{P,R}(k^2)$
and $\Phi(k^2)$, together with $\bar s$ and $s_0$ which separates high
and low energy parts of the quark spectral density
(\ref{spectrum},\ref{spectrum:I}). These parameters are determined by
fits to the nucleon structure functions at large $Q^2$. In our analysis
we use recent NMC data \cite{NMColl92} for $F_2(x)$ and neutrino
data \cite{Mattis90,CDHSW91} for $F_2^N(x)$, $xF_3^N(x)$ and $\bar q(x)$ at
average momentum transfer $Q^2\approx 5-10\,$GeV$^2$. Our result shown
in Fig.~\ref{fig4} uses
\begin{eqnarray}
\Lambda_V^2=1.2\,\mbox{GeV}^2, \nonumber\\
\label{cutoffs}
\Lambda_P^2=2.5\,\mbox{GeV}^2, \\
\Lambda_R^2=4.0\,\mbox{GeV}^2, \nonumber
\end{eqnarray}
\vspace*{-\belowdisplayskip}%
together with \vspace*{-\baselineskip}\vspace*{-\abovedisplayshortskip}
\begin{equation}  \label{s0}
s_0 = 2\bar s = 4\,\mbox{GeV}^2.
\end{equation}

\vspace*{-\belowdisplayshortskip}%
The global view in Fig.~\ref{fig4} is supplemented by more detailed
comparisons with data at large $x$ (Fig.~\ref{fig5}a) and at small $x$
(Fig.~\ref{fig5}b). In particular, Fig.~\ref{fig5}a demonstrates that the
exponents $n_V=n_P=n_R=4$ are properly chosen to reproduce tails of the
distribution functions, while Fig.~\ref{fig5}b focuses on the behavior at
$x<0.1$ where sea quarks begin to dominate $F_2^N$.

A remark concerning the value of $\bar s$ is in order.  This parameter
roughly corresponds to the average squared mass of the residual quark-gluon
system, with one quark removed from the nucleon. The value of $\bar s$
depends on the scale at which we study the system. For example, in the bag
model (or in a constituent quark model) $\sqrt{\bar s_{\rm bag}}
\sim\frac23 M$, where $M$ is the nucleon mass. This corresponds to the
quark distributions at a low resolution scale $Q^2 < 1\,$GeV$^2$, which
then has to be evolved to the momentum transfer $Q^2$ at which the
experimental data are taken (e.g.\cite{ScSiTh91}).  In terms of the quark
spectral densities (\ref{rho-R},\ref{rho-L}) the $Q^2$-evolution effect
modifies the spectrum of intermediate states by adding radiative
corrections which generate gluons and $q\bar q$ pairs. This shifts the
quark spectra to larger values of $s$. In our approach we fit to the
measured structure functions at large $Q^2$ and effectively incorporate the
$Q^2$-evolution effect in the parameter $\bar s$, averaged over the $Q^2$
range of the experimental data.

\subsection{Structure function at small $Q^2$ and small $x$}

So far we have only discussed the scaling region, $Q^2>5\,$GeV$^2$. New
phenomena enter at small $Q^2$ and small $x$. This is also where shadowing
effects show up in the nuclear structure functions $F_2^A$, so that special
attention is assigned to this region.

At small $Q^2 < 1\,\mbox{GeV}^2$ the scaling behavior is violated.
Furthermore conservation of the electromagnetic current requires that $F_2$
must vanish as $Q^2$ goes to zero. We therefore need a model which provides
a smooth transition from the scaling to non-scaling regions. Here we
discuss a model based on the generalized vector meson dominance (GVMD)
ideas \cite{PilWei90}.
In the GVMD approach the structure function $F_2$ is expressed in terms
of a dimensionless spectral function $\Pi(\mu^2)$ of hadronic states which
couple to the photon:
\begin{equation}  \label{GVMD}
F_2^N(x,Q^2) = \frac{Q^2}{\pi}\int_0^{\infty}d\mu^2\,
\frac{\mu^2\Pi(\mu^2)}{(\mu^2+Q^2)^2}\, \sigma_N(\mu^2,S).
\end{equation}
Here $\sigma_N(\mu^2,S)$ is the cross section for scattering of a hadronic
state with mass $\mu$ from the nucleon, and $S=M^2 + Q^2(1/x-1)$ is the
total squared center-of-mass energy of the virtual photon-nucleon system.
The distribution $\Pi$ represents the spectrum of correlated
quark-antiquark pairs and multi-meson states with spin and parity $J^{\pi}
= 1^-$. It separates into a low-mass part dominated by the vector mesons
and a high-mass continuum part $\Pi_c$:
\begin{equation}  \label{Pi}
\Pi(\mu^2) = \sum\limits_{V=\rho,\omega,\phi}
          \left(\frac{m_V^2}{g_V^2}\right)\,\delta(\mu^2 - m_V^2) +
          \Pi_c(\mu^2)\theta(\mu^2 - \mu_0^2),
\end{equation}
where the continuum starts at $\mu_0\gtrsim 1\,$GeV, just above the $\phi$
meson mass. For the vector meson masses $m_V$, their coupling constants
$g_V$ and cross sections $\sigma_{\rm VN}$ we use standard values
summarized in Table~I:

\vspace*{0.5cm}
\begin{center}
\begin{tabular}{|c|c|c|c|}\hline
 $\;V\;$  &  $\;m_V\;$(MeV) & $\;g_V^2/4\pi\;$ & $\;\sigma_{\rm VN}\;$(mb)\\
\hline
 $\rho   $&    768.3   &    2.38    &    27              \\ \hline
 $\omega $&    782.0   &   18.4     &    27              \\ \hline
 $\phi   $&    1019.4  &   13.8     &    12              \\ \hline
\end{tabular}
\end{center}

\begin{center}
\begin{minipage}{12cm}
 Table I: Vector meson masses $m_V$ \cite{PADAGR90}, coupling constants
          $g_V$  \cite{SchStr76} and cross sections $\sigma_{\rm VN}$
          \cite{BaSpYe78}.
\end{minipage}
\end{center}
\vspace*{0.5cm}

\noindent
The vector meson part is then
\begin{equation}  \label{VM}
F_2^{\rm N,VM} = \frac{Q^2}{\pi}\sum\limits_{V=\rho,\omega,\phi}
          \left(\frac{m_V^2}{g_V}\right)^2\,
           \left(\frac{1}{m_V^2+Q^2}\right)^2\,
                  \sigma_{\rm VN}.
\end{equation}
It dominates the structure function at small $Q^2 < 1\,\mbox{GeV}^2$.  On
the other hand, at large momentum transfer $Q^2\gg m_V^2$ the vector-meson
contribution (\ref{VM}) vanishes as $\sim m_V^2/Q^2$. In this region large
masses $\mu^2\sim Q^2$ in the $\Pi_c$ part of the spectral function
(\ref{Pi}) take over.  At large $Q^2 \gg \mu_0^2$ this piece leads to a
structure function with proper scaling behavior.

With this consideration in mind, we make the following ansatz in order to
interpolate between the regions of small and large $Q^2$:
\footnote{Eq.(\ref{F2Ntotal}) is similar to a
structure function model discussed in \cite{BadKwi92a}.
}
\begin{equation}   \label{F2Ntotal}
F_2^N(x,Q^2) = F_2^{\rm N,VM}(x,Q^2) +
               \frac{Q^2}{Q^2+Q_0^2} \,F_2^{\rm N,as}(x).
\end{equation}
For the asymptotic part $F_2^{N,as}(x)$ of the structure function we use
our model as described in the previous section. The parameter $Q_0^2$ is
expected to be of order $\mu_0^2 \sim 1\,$ GeV$^2$.
Scaling occurs for $Q^2 \gg Q_0^2$.

In Fig.~\ref{fig6} we show the characteristic behavior of $F_2^N(x,Q^2)$ at
a small value of the Bjorken variable ($x=0.01$) as a function of $Q^2$.
We find that $Q_0^2=2\,$GeV$^2$ gives a good description of the data.  One
observes that vector mesons dominate at $Q^2\ll 1\,$GeV$^2$, whereas the
scattering of uncorrelated $q\bar q$ pairs governs $F_2^N$ for $Q^2 >
2\,$GeV$^2$. At $Q^2=1\,$GeV$^2$, roughly one third of $F_2^N$ at $x=0.01$
is due to vector mesons; at $Q^2=5\,$GeV$^2$ they still contribute about
10\%.  This is in agreement with results obtained within the framework of
the generalized vector meson dominance model \cite{PilWei90}.

\section{Nuclear Structure Functions}
\label{kern}

We now turn to our main theme, namely DIS on nuclear targets and the
structure function $F_2^A(x,Q^2)$.

Let us first discuss the region of small $x$. We have pointed out that deep
inelastic scattering at small $x$ (as seen from the target rest frame)
proceeds dominantly via the mechanism (II), and we now investigate nuclear
effects based on this observation. As we already mentioned, for $x<0.1$ the
propagation length $\lambda\sim (Mx)^{-1}$ of $q\bar q$-fluctuations in the
photon wave function exceeds the average distance $d \sim 1.8$\,fm between
bound nucleons in the nucleus. The propagating $q\bar q$ pair can then
interact coherently with several nucleons. This multiple scattering effect
becomes significant when $\lambda$ is larger than the quark mean free path
$l=(\rho\sigma_{qN})^{-1}$, where $\rho$ is the average nuclear density. In
terms of the $x$ variable the last condition reads $x < \rho\sigma_{qN}/M$.
The nuclear effects due to multiple scattering of the $q\bar q$-pair will
saturate for $x < (MR_A)^{-1}$, i.e. when the propagation length $\lambda$
exceeds the nuclear radius $R_A$. In this case the virtual photon converts
into a $q\bar q$ pair already outside the nucleus. This pair interacts with
nucleons at the nuclear surface which absorbs part of the incoming flux and
thereby screens the inner nucleons.  This is the {\it shadowing} effect
which we study in detail in the next section.

On the other hand coherent multiple scattering effects are not important at
large $x$ where $\lambda < d$ and the $q\bar q$ fluctuation has no time to
scatter more then once. In this region the DIS process takes place mainly
on a single nucleon in the nucleus. Effects due to nuclear binding and
Fermi-motion are now important. We discuss these in detail in
Section~\ref{largex}.

\subsection{The small $x$ region: Shadowing}
\label{smallx}

As in our previous discussion of the free nucleon structure function $F_2^N$,
we use the following ansatz for the nuclear structure function:
\begin{equation}   \label{F2Atotal}
F_2^A(x,Q^2) = F_2^{\rm A,VM}(x,Q^2) +
               \frac{Q^2}{Q^2+Q_0^2} \,F_2^{\rm A,as}(x),
\end{equation}
and discuss the asymptotic (scaling) and vector meson contributions to
$F_2^A$ separately.

Consider first the scaling part $F_2^{\rm A, as}(x)$.  It is formally
obtained from $F_2^N(x)$ in our model through the replacement of the
quark-nucleon and antiquark-nucleon amplitudes $T_{qN}$ and $T_{\bar qN}$
in eqs.(\ref{qqbarII}) by corresponding nuclear amplitudes, $T_{qA}$ and
$T_{\bar qA}$. For the type (I) distributions in eq.(\ref{sum-q}) we simply
use $q_A^{\rm (I)}(x)=A q_N^{\rm (I)}(x)$; the validity of this
approximation will be discussed in the next section.  We use Glauber
multiple scattering theory \cite{Glaube59} to express the nuclear
amplitudes $T_A$ in terms of the $T_N$:
\begin{equation} \label{TA}
T_A(s,t)=-2is\int d^2 {\bf b}\: e^{i{\bf q'\cdot b}}
\langle
A|\exp\left(i\sum_j\chi({\bf b}-{\bf b}_j)\right)-1|A
\rangle.
\end{equation}
Here the averaging is performed over the nuclear wave function and the
sum runs over all bound nucleons located at positions ${\bf r}_j=({\bf b}_j,
z_j)$ in the nuclear c.m. frame; ${\bf q}'$ is the momentum transfer,
$t=-{\bf q}'^2$. The eikonal phase $\chi({\bf b})$ is related to the nucleon
amplitude $T_N$ as follows:
\begin{equation} \label{TN}
T_N(s,t)=-2is\int d^2{\bf b}\: e^{i{\bf q'\cdot b}}
\left(\exp(i\chi({\bf b}))-1\right) .
\end{equation}
The Glauber analysis of high-energy hadron scattering from nuclei
indicates that the result for $T_A$ is not sensitive to $NN$-correlations and
other details of the nuclear wave function \cite{Franco72}. We can therefore
evaluate the nuclear matrix element in eq.(\ref{TA}) in a simple
approximation assuming that the squared nuclear wave function is given by the
product of Gaussian one-particle densities:
\begin{equation} \label{wf}
\left|\Psi_A({\bf r}_1,\ldots,{\bf r}_A)\right|^2=
\prod_{j=1}^A(\pi R^2)^{-3/2}\exp(-r_j^2/R^2) .
\end{equation}
The parameter $R^2=\frac23 R_A^2$ is fixed by the nuclear root mean square
radius $R_A=1.12\,A^{1/3}\:$fm as determined by electron scattering data.
Using this wave function one can easily calculate $T_A$ in terms of $T_N$,
with the result:
\begin{equation} \label{TGauss}
T_A(s,k^2)=T_N(s,k^2)\sum_{j=1}^A\frac{1}{j}{A \choose j}
\left(\frac{iT_N(s,k^2)}{2\pi s R^2}\right)^{j-1} .
\end{equation}
The effective number $n_{\rm eff}$ of terms which contribute to the sum
(\ref{TGauss}) can be estimated as the average number of rescatterings of a
classical particle moving along the diameter of the nucleus:  $n_{\rm eff}
= 2R/l$ where $l$ is the quark mean free path, and $l\approx3$~fm for a
quark-nucleon cross section $\sigma_{qN}\approx13\:$mb.  Therefore the
triple scattering term $j=3$ practically saturates the multiple scattering
series for nuclei up to $A\sim 100$. In the actual calculations we keep
four terms in eq.(\ref{TGauss}).

The calculated ratios $F_2^A(x)/AF_2^N(x)$ are shown as dashed curves in
Fig.~\ref{fig7} for several nuclei. Comparing these results with the recent
NMC data \cite{NMColl91}, we find that roughly only half of the measured
shadowing effect can be explained in this way. In other words, mechanism
(II) alone, with quark- and antiquark-nucleon interactions constrained by
high energy $pp$ and $p\bar p$ data, cannot account for all of the observed
nuclear shadowing.

We find this not surprising, for the following reason. The experimental
data at small $x$ are taken at relatively small values of $Q^2$. For
example at $x\approx 0.01$ the average momentum transfer is $Q^2\approx
1.6\,$GeV$^2$. But at these low $Q^2$ it is not justified to consider only
leading twist contributions to the structure function. The multiple
scattering of strongly correlated $q\bar q$ pairs on the nuclear target now
becomes important.  This brings in the vector meson contribution
$F_2^{A,VM}$ to the nuclear structure function (\ref{F2Atotal}). Its form
is analogous to that of $F_2^{\rm N,VM}$ for the free nucleon (see
eq.(\ref{VM})), where now the vector meson-nucleon cross sections
$\sigma_{\rm VN}$ are replaced by corresponding nuclear cross sections
$\sigma_{\rm VA}$. The latter is related to the former via the Glauber
multiple scattering series (\ref{TGauss}). Together with the scaling part
$F_2^{\rm A,as}(x)$, which includes the nuclear interaction of uncorrelated
$q\bar q$-pairs, we then calculate the full nuclear structure function
$F_2^A(x,Q^2)$ according to eq.(\ref{F2Atotal}).

In Fig.~\ref{fig7} we compare our results for $R(x,Q^2)
=F_2^A(x,Q^2)/AF_2^N(x,Q^2)$, including vector mesons, with the data of the
NMC collaboration \cite{NMColl91}. For every $x$-bin, $R(x,Q^2)$ is
calculated using the corresponding average $\left< Q^2 \right>$ given in
\cite{NMColl91}. We see that the vector mesons can account quite well for
the missing part of the shadowing.

In summary we find that the scaling contribution to the nucleon structure
function alone can account for only about half of the measured shadowing
effect. The other half results from the interactions of strongly correlated
$q\bar q$ pairs, i.e. vector mesons, with the target. The fact that the
observed shadowing is only weakly $Q^2$ dependent \cite{NMColl91} has now a
plausible explanation: although the vector meson contributions vanish at
large $Q^2$ there is still a sizable shadowing effect due to the
interaction of uncorrelated quarks or antiquarks with the nuclear target.

\subsection{The large $x$ region: Convolution model}
\label{largex}

In the region $x>0.2$ nuclear structure functions are commonly described
within the impulse approximation (see Fig.~\ref{fig8}) ignoring final state
interaction of the nucleon debris with remaining nuclear system.%
\footnote{An attempt to estimate the influence of final state interaction on
leading twist nuclear structure functions was  made
in ref. \cite{ThMiSc89,SaiTho93}.}
The Feynman  diagram in Fig.~\ref{fig8} is usually written in the form of a
convolution (see e.g. \cite{Jaffe85}):
\begin{equation}    \label{convol:naive}
F_2^A(x)=\int_{x}\!dy\; D_{\rm N/A}(y)\:F_2^N(x/y) .
\end{equation}
Here the structure function $F_2^N(x)$ is folded with the (light-cone)
momentum distribution $D_{\rm N/A}(y)$ of nucleons in the nucleus.
Eq.(\ref{convol:naive}) has frequently been applied in calculations of
Fermi-motion and binding corrections
\cite{AkKuVa85,BiLeSh89,DunTho86,Kulagi89,JunMil88,CioLiu89,DieMil91}.
The convolution formalism is also used to evaluate meson cloud effects
\cite{MuScMe92}, exchange currents corrections \cite{Kulagi89,EriTho83}
etc. (for a review see ref.\cite{BicTho89}).

However, as pointed out in a recent analysis \cite{MeScTh93}, a derivation
of the convolution model (\ref{convol:naive}) even within the impulse
approximation implies several assumptions which are generally not
justified. In this section we re-examine the derivation of
eq.(\ref{convol:naive}) on the basis of our covariant approach developed in
Section~\ref{frame}. It will turn out that off-shell effects are important,
so that the simple convolution model is generally not a good approximation
for $F_2^A$.

We start from eq.(\ref{f}) which gives the nuclear light-cone distribution
function $f_A(x_A)$ in terms of the quark correlator $\Delta_A(k,P_A)$ in
the nucleus.  Consider now the diagram Fig.~\ref{fig8}. It can be written
as a convolution of the quark correlator $\widehat\Delta_N(k,p)$ in the
bound nucleon and the nucleon propagator $G(p,P_A)$ in the nucleus:
\begin{equation}
\label{convol:D}
\Delta_A(k,P_A)=-i \int\! \frac{d^4p}{(2\pi)^4}\: {\rm Tr}_N
\left(\widehat\Delta_N(k,p)G(p,P_A)\right) .
\end{equation}
Here and in the following the ``hat" on $\widehat\Delta_N$ and other
quantities indicates their matrix structure in  nucleon  Dirac space,
and the trace Tr$_N$ is taken with respect to nucleon variables.  The
four-momenta $P_A$, $p$ and $k$ refer  to the nucleus, the bound nucleon
and the quark in the nucleon, respectively. The quark propagator in the
on-mass-shell nucleon (\ref{Delta}) averaged over nucleon polarizations is
related to $\widehat\Delta_N$ as follows:
\begin{equation}  \label{D-av}
\Delta_N(k,p)=\frac12
{\rm Tr}_N\left(\widehat\Delta_N(k,p)(\overlay{\slash}p+M)\right) .
\end{equation}
Substituting eq.(\ref{convol:D}) into eq.(\ref{f}) we obtain for the
nuclear light-cone distribution,
\begin{equation}
\label{convol:fA}
f_A(x_A) = -i \int\! \frac{d^4p}{(2\pi)^4} {\rm Tr}_N
\left(\widehat f_N(x',p^2)\,G(p,P_A)\right) ,
\end{equation}
with $x_A=Q^2/2P_A\!\cdot\!q$ and
\begin{equation}
\label{f-hat}
\widehat f_N(x',p^2) = -i \int\! \frac{d^4k}{(2\pi)^4}\:
    \frac{{\rm Tr}_Q\left(\overlay{\slash}q\,\widehat\Delta_N(k,p)\right)}
            {2p\!\cdot\!q}
        \:\delta\!\left(x'-\frac{k\!\cdot\!q}{p\!\cdot\!q}\right).
\end{equation}
Here the trace Tr$_Q$ is taken with respect to quark variables and
$x'=Q^2/2p\!\cdot\!q$ is the Bjorken variable of an off-shell nucleon
with four-momentum $p$.

Let us now examine the Lorentz structure of $\widehat f_N$ in the nucleon
Dirac space. In general it can be expanded in terms of a complete set of 16
Dirac matrices. However there are only four independent terms%
\footnote{
We ignore possible terms proportional to $\gamma_5$ and
$\gamma_{\alpha}\gamma_5$ which do not
contribute to the unpolarized structure functions.
}
which can be constructed from the momenta $p$, $q$ and the Dirac matrices:
\begin{eqnarray}    \label{f-hat-Dirac}
\widehat f_N=f_S\,I + f^{\mu}_V\,\gamma_{\mu} +
        f^{\alpha\beta}_T\,\sigma_{\alpha\beta},
\end{eqnarray}
with
\begin{mathletters} \label{f-i}
\begin{eqnarray}
f_S &=& {f_0\over 2M},\\
f^{\mu}_V &=& \frac{f_1}{2M^2}\, p^{\mu}
        + \frac{f_2}{2p\!\cdot\!q}\, q^{\mu},\\
f^{\alpha\beta}_T &=& \frac{f_3}{2p\!\cdot\!q\,M}\, p^{\alpha}q^{\beta} ,
\end{eqnarray}
\end{mathletters}%
where $\sigma_{\alpha \beta}=\frac12[\gamma_{\alpha},\gamma_{\beta}]$ and
$f_i(x,p^2),\ i=0,1,2,3,$ are dimensionless Lorentz invariant functions.
The coefficients in eq.(\ref{f-i}) are chosen in such a way that the
on-mass-shell distribution function averaged over the nucleon spin is given
by:
\begin{equation}    \label{fN-fi}
f_N(x)=\lim_{p^2\to M^2}\frac12{\rm Tr}
        [(\overlay{\slash}p+M)\widehat f_N(x,p^2)]
            = f_0+f_1+f_2 .
\end{equation}
We see that the tensor term in (\ref{f-hat-Dirac}) does not contribute to
the unpolarized structure functions.

Substituting (\ref{f-hat-Dirac}) into (\ref{convol:fA}) one finds an
equation which connects the nuclear light-cone distribution with the
functions $f_i$:
\begin{equation}    \label{fA-fi}
f_A(x_A)=-i\int\! \frac{d^4p}{(2\pi)^4}
\sum_{i=0}^3 {\cal C}_i(p,q)\,f_i(x',p^2) .
\end{equation}
The functions ${\cal C}_i(p,q)$ are given by traces of the nucleon
propagator $G$ with the different Dirac matrices in eq.(\ref{f-hat-Dirac}):
\begin{mathletters} \label{C-i}
\begin{eqnarray}
{\cal C}_0 &=& {1\over 2M}\,{\rm Tr}\, G(p,P_A), \\
{\cal C}_1 &=& {1\over 2M^2}\,{\rm Tr}\,(G(p,P_A)\overlay{\slash}p), \\
{\cal C}_2 &=& \frac1{2p\!\cdot\!q}{\rm Tr}\,(G(p,P_A)\overlay{\slash}q),\\
{\cal C}_3 &=& {p^{\alpha}q^{\beta}\over 2p\!\cdot\!qM}\,
{\rm Tr}\left(G(p,P_A)\sigma_{\alpha\beta}\right) .
\end{eqnarray}
\end{mathletters}%

We emphasize here that in general eq.(\ref{fA-fi}) does not reduce to the
simple  convolution formula (\ref{convol:naive}) with respect to the
light-cone momentum. There are two reasons for this.  First, terms with
different Lorentz structures in eq.(\ref{f-hat-Dirac}) are convoluted with
correspondingly different nuclear distribution functions.  Secondly, the
structure functions in the off-shell region depend not only on the scaling
variable $x'$ but also on the off-shell mass $p^2$ of the bound nucleon.
This will now be examined in more detail.

In the following we shall consider the nuclear structure functions as
functions of the ``nucleon" Bjorken variable $x=Q^2/2Mq_0$ instead of the
``nuclear" one, $x_A=Q^2/2M_Aq_0$. The structure function $F_2^A$ as a
function $x$ reads:
\begin{eqnarray}
F_2^A(x) &=& x\left({\cal F}_A(x)-{\cal F}_A(-x)\right),\\
{\cal F}_A(x) &=& \frac{M}{M_A}f_A(\frac{M}{M_A}x)\label{fA(x)} .
\end{eqnarray}
One can easily see that the transformation (\ref{fA(x)}) preserves the
normalization of the nuclear distribution function as a function
of $x$:
\begin{equation}    \label{3A}
\int\limits_{-M_A/M}^{M_A/M}\!\!\! dx\,{\cal F}_A(x)=
\int\limits_{-1}^{1}\! dx_A f_A(x_A) = 3A .
\end{equation}
Eq.(\ref{fA-fi}) can be simplified if one assumes that the nucleus is a
nonrelativistic system. In this case, as shown in the Appendix, ${\cal A}_3$
vanishes up to terms of order $|{\bf p}|^3/M^3$ if one uses the
nonrelativistic form for the nucleon propagator.  Moreover ${\cal
A}_3=0$ for spinless nuclei. In the same approximation the functions ${\cal
A}_0$, ${\cal A}_1$ and ${\cal A}_2$ are proportional to each other (see
eqs.(\ref{trs}) in the Appendix).  This allows us to introduce one {\it
unique} nucleon distribution function $D_{\rm N/A}$ which however depends
also on $p^2$. Using eqs.(\ref{trs}) in the Appendix we have,
\begin{eqnarray}
\label{convol:nr}
{\cal F}_A(x) &=&
\int\limits_{y>|x|}\!\frac{dy}y\int dp^2\,D_{N/A}(y,p^2)f_N(x/y,p^2),\\
\label{f-offshell}
f_N(x,p^2) &=& \sqrt{p^2/M^2}f_0(x,p^2) +
        (p^2/M^2)f_1(x,p^2) + f_2(x,p^2),\\
\label{D-N/A}
D_{\text{N/A}}(y,p^2) &=& \int\!\frac{d^4p'}{(2\pi)^4}\, S(p')
        \left(1+\frac{p'_3}{M}\right)
\delta(y -\frac{p'_+}{M})\,\delta(p^2 - p'^2).
\end{eqnarray}
Eq.(\ref{f-offshell}) can be identified with the light-cone distribution
function of the bound nucleon. In eq.(\ref{D-N/A}) we have introduced
the nuclear spectral function
\begin{equation}  \label{S(p)}
    S(p)=2\pi\sum_n\delta(p_0-M-\varepsilon_n)
    \left|\Psi_n({\bf p})\right|^2,
\end{equation}
where the sum includes all residual nuclear states with $A-1$ nucleons
which carry together the momentum $-{\bf p}$. All other quantum numbers are
denoted as $n$.  The nucleon separation energy is defined as
$\varepsilon_n=E_0(A) - E_n(A-1)$. Furthermore $\Psi_n({\bf
p})=\langle(A-1)_n,-{\bf p}|\Psi(0)|A\rangle$ where $\Psi(0)$ is the
nonrelativistic nucleon field operator at ${\bf r}=0$.  The spectral
function is normalized to the number of nucleons in the nucleus,
\begin{equation}
\label{Snorm}
\int\!\frac{d^4p}{(2\pi)^4}\, S(p)=A,
\end{equation}
which guarantees the correct normalization of the nucleon distribution
function in eq.(\ref{D-N/A}).

It is convenient to introduce also the distributions of quarks, $q(x,p^2)$,
and antiquarks, $\bar q(x,p^2)$, in the bound nucleon. These are expressed
through $f_N(x,p^2)$ by eqs.(\ref{qqbar}):
\begin{mathletters}
\begin{eqnarray}
q(x,p^2)      &=& f_N(x,p^2),\\
\bar q(x,p^2) &=& -f_N(-x,p^2)
\end{eqnarray}
\end{mathletters}%
In terms of these distributions the structure functions of the bound nucleon
are given by the usual parton model formula (see eq.(\ref{F2IA})).
The relation between the nuclear and the bound nucleon structure functions
then reads:
\begin{equation}    \label{F2A}
F_2^A(x)=\int\limits_x^{M_A/M}\!\!\!\! dy\int\!dp^2\,
D_{\rm N/A}(y,p^2) F_2^N(x/y;p^2).
\end{equation}
Let us examine this equation  in more detail. The nucleon distribution
(\ref{D-N/A}) is strongly peaked around $p^2=M^2$ and $y=1$,  with
a characteristic width $\Delta y\sim p_F/M$, where $p_F$ is the nucleon
Fermi-momentum.  Expanding the bound nucleon structure function  in
eq.(\ref{convol:nr}) in a Taylor series around these points and integrating
term by term, one then obtains the following expression for the nuclear
structure function per nucleon:
\begin{eqnarray}    \label{expansion}
F^A_2(x)/A \simeq F_2^N(x)
     -\frac{\langle\varepsilon\rangle}{M}\:x{F_2^N}^{\prime}(x)
     +\frac{\langle T\rangle}{3M}\:x^2{F_2^N}^{\prime\prime}(x) \nonumber\\
     +2\:\frac{\langle\varepsilon\rangle-\langle T\rangle}{M}
\left(p^2\frac{\partial F_2^N(x;p^2)}{\partial p^2}\right)_{p^2=M^2} .
\end{eqnarray}
Here ${F_2^N}^{\prime}(x)$ and ${F_2^N}^{\prime\prime}(x)$ are derivatives
of the structure function with respect to $x$, and
$\langle\varepsilon\rangle$ and $\langle T\rangle$ are the mean separation
and kinetic energies of the bound nucleon,
\begin{eqnarray}    \label{E-rem}
    \langle\varepsilon\rangle &=&
    {1\over A}\int\!\frac{d^4p}{(2\pi)^4}\, S(p)\, \varepsilon,\\
\label{E-kin}
\langle T\rangle &=&
{1\over A}\int\!\frac{d^4p}{(2\pi)^4}\, S(p)\, \frac{{\bf p}^2}{2M}.
\end{eqnarray}
Corrections to eq.(\ref{expansion}) are of higher order in
$\langle\varepsilon\rangle/M$ and $\langle T\rangle/M$. One should also
note that eq.(\ref{expansion}) can safely be used for $1-x > p_F/M \sim
0.3$.  In this region the condition $x/y\le 1$ gives practically no
restrictions on the integration over the nucleon momentum $p$ in
eqs.(\ref{E-rem},\ref{E-kin}).

The first three terms in eq.(\ref{expansion}) are identical to the result
of \cite{AkKuVa85,Kulagi89} in their discussion of the EMC-effect, while
the last one reflects the leading contribution from the $p^2$-dependence of
the bound nucleon structure function. Let us first neglect the latter and
discuss effects due to separation and kinetic energies.  These terms are of
opposite signs and the competition between them results in a behavior of
the ratio $R(x)=F_2^A/A F_2^N$ at  $x>0.3$ similar to that seen in the
experiment.  As it was first pointed out in ref.\cite{AkKuVa85} and
discussed by many authors (see e.g. ref.\cite{BicTho89} for a review), this
may account for the EMC-effect at intermediate and large $x$.  An important
(and still open) problem in this respect is a reliable calculation of
$\langle\varepsilon\rangle$ and $\langle T\rangle$.  In a simple nuclear
shell model the removal energy is averaged over all occupied levels.  One
finds typical values $\langle\varepsilon\rangle\approx -(20-25)\;$MeV and
$\langle T\rangle\approx 18-20\;$MeV.  Correlations between nucleons change
the simple mean field picture significantly and  lead to high momentum ($p
> p_F$) components in the nuclear spectral function (\ref{S(p)}). This
in turn causes an increase of the average removal energy
$\langle\varepsilon\rangle$.  In order to demonstrate this let us consider
the Koltun sum rule \cite{Koltun72}, which is exact if only two-body forces
are present in the nuclear Hamiltonian:
\begin{equation} \label{Kolsr}
\langle\varepsilon\rangle + \langle T\rangle = 2\mu_B .
\end{equation}
Here $\mu_B\approx - 8\,$MeV is the nuclear binding energy per
nucleon. In particular, eq.(\ref{Kolsr}) tells that an increase  of
$\langle T\rangle$ due to high momentum components implies also an
increase of $|\langle\varepsilon\rangle|$. We refer in this
respect to a recent calculation \cite{BeFaFa89} of the spectral function
of nuclear matter based on a variational method.  This calculation
shows that there is a significant probability to find nucleons with
high momentum and large separation energies. Integration of the
spectral function of ref.\cite{BeFaFa89} gives $\langle
T\rangle\approx 38\;$MeV and $\langle\varepsilon\rangle\approx
-70\;$MeV.  In order to estimate these quantities for finite nuclei
one usually assumes \cite{CioLiu89} that the high momentum component
of the nucleon momentum distribution is about the same as in nuclear
matter, which gives $\langle T\rangle\approx 35\;$MeV for a wide range
of nuclei.  The latter quantity together with eq.(\ref{Kolsr}) leads
to $\langle\varepsilon\rangle\approx -50\;$MeV.  It should be noted
however that, even though a qualitative understanding of the EMC effect can
be obtained using such values for $\langle\varepsilon\rangle$ and
$\langle T\rangle$, a quantitative description is still lacking.

Let us finally discuss nuclear effects due to the off-mass shell properties
of nucleons bound in  nuclei. We note in this respect that the analysis of
section \ref{frame} can be applied also to eq.(\ref{f-hat}) which describes
the quark and the antiquark distributions of the off-shell nucleon. We
parameterize the loop momentum in eq.(\ref{f-hat}) in terms of the Sudakov
variables, $k=\alpha p+\beta q'+k_{\perp}$, where $p$ is the nucleon
momentum in eq.(\ref{convol:D}) and $q'=q+x'p$. Assuming again that the
quark correlator $\widehat\Delta_N$ is an analytic function of $s=(p-k)^2,\
u=(p+k)^2$ and $k^2$ we end up with equations similar to
eqs.(\ref{qqbar:DI},\ref{ksqr}), where however the squared nucleon mass
$M^2$ is now replaced by $p^2$. In particular, for the quark distribution
$q(x',p^2)$ we have
\begin{equation}  \label{q-offshell}
q(x',p^2)=\frac1{16\pi^2}\int ds\int_{-\infty}^{k_{\rm max}^2}\! dk^2
           \:\rho_R(s,k^2,x',p^2),
\end{equation}
where $k_{\rm max}^2=x'\left(s/(x'-1)\!+\!p^2\right)$ (cf. eq.(\ref{ksqr}))
and $\rho_R$ is the quark spectral density given by an equation similar to
eq.(\ref{rho}).

We conclude from eq.(\ref{q-offshell}) that the $p^2$-dependence of the
structure functions has two primary sources:
\begin{enumerate}
\item[1)]
Explicit dependence of the quark correlator  $\widehat\Delta_N$ in
eq.(\ref{f-hat}) on $p^2$ (``dynamical'' $p^2$-dependence). This leads
to a $p^2$-dependence of parameters characterizing the quark spectral density
$\rho_R$.
\item[2)]
Dependence of the invariant variables $s$ and $u$ on $p^2$ (``kinematical''
$p^2$-dependence). This manifests itself via the $p^2$ dependence of $k_{\rm
max}^2$  in eq.(\ref{q-offshell}).
\end{enumerate}

In this section we investigate nuclear effects at $x>0.2$.  Therefore we
keep contributions to the quark spectral density $\rho_R$ only from
mechanism (I) which dominates in the large $x$ region. Recalling
eq.(\ref{spectrum:I}) we now have
\begin{equation}
\rho_R(s,k^2,x',p^2)=\Phi(k^2,p^2)\delta(s-\bar s),
\end{equation}
and evaluate the derivative of the bound nucleon structure function with
respect to $p^2$:
\begin{eqnarray} \label{dFdp2}
{\partial F_2^N(x,p^2)\over \partial p^2} \approx
x{\partial q^{(I)}(x,p^2)\over \partial p^2} = \nonumber\\
{x \over 16\pi^2}
\left(
    x\Phi(k_{{\rm max}}^2,p^2)
+ \int_{-\infty}^{k_{{\rm max}}^2 }\! dk^2\:
    \frac{\partial\Phi(k^2,p^2)}{\partial p^2}
\right).
\end{eqnarray}
The two terms in eq.(\ref{dFdp2}) reflect the two sources of the
$p^2$-dependence. The first one arises from the $p^2$-dependence of $k_{\rm
max}^2$. Its contribution to $F_2^A$ in eq.(\ref{expansion}) leads to an
enhancement of the binding effect (see Fig.~\ref{fig9}).  However if only
this ``kinematical'' $p^2$-dependence would be present the number of
valence quarks in the nucleon would  change with $p^2$.  Therefore an
explicit $p^2$-dependence of the quark spectral density is necessary to fix
the normalization of the valence quark distribution.  In order to evaluate
this effect we assume that the $k^2$-dependence of the function
$\Phi(k^2,p^2)$ is the same as for the on-mass-shell nucleon,
eq.(\ref{g(k2)}), and that the $p^2$-dependence comes from the
corresponding dependence of the cut-off parameter $\Lambda_V$,
$\Phi(k^2,p^2)=\Phi(k^2;\Lambda_V(p^2))$. We fix the $p^2$-dependence of
$\Lambda_V$ in such a way that the number of valence quarks in the nucleon
is independent of the off-shell nucleon mass $p^2$, i.e.
\begin{equation}
\frac{\partial }{\partial p^2} \int_0^1\!\! dx\: q^{(I)}(x,p^2) = 0.
\end{equation}
Together with eq.(\ref{dFdp2}) we can now calculate $\partial
\Lambda_V^2/\partial p^2$.  We find $\partial \Lambda_V^2/\partial
p^2\approx 0.15$ at $p^2\!=\!M^2$.

The $p^2$ dependence of $\Lambda_V$ can be viewed as an effect of changing
the bound nucleon size $R_N\sim 1/\Lambda_V$ in nuclei. In fact
$\partial \Lambda_V^2/\partial p^2 > 0$ implies a ``swelling" of the nucleon
in the nuclear environment.  For the relative change of the radius of the
bound nucleon we obtain,
\begin{equation}
\label{dR}
\frac{\delta R_N}{R_N}\sim -\frac12\frac{\delta\Lambda_V^2}{\Lambda_V^2}=
-\frac{\partial \Lambda_V^2}{\partial p^2}\,
  \frac{M \langle\varepsilon  -  T\rangle}{\Lambda_V^2}.
\end{equation}
In this way we find a very small ($0.8$\%) increase of the size of a nucleon
bound in the nucleus, indicating a remarkable stability of the nucleon.%
\footnote{
In ref.\cite{BGNiPZ93} the EMC effect is attributed to
a 10\% increase of the nucleon radius in the nucleus.
We emphasize here that in our study of nuclear binding
and off-shell effects this increase of size is reduced
by an order of magnitude.
}
This result is compatible with recent findings using quite a different
approach \cite{SaiTho93}.

In Fig.~\ref{fig9} we show  a typical result for $R(x) = F_2^A(x)/(A\,
F_2^N(x))$ calculated using eq.(\ref{expansion}).  The solid curve
corresponds to the impulse approximation neglecting any $p^2$-dependence.
The dashed curve is the result of the full calculation including
eq.(\ref{dFdp2}). The ``kinematical'' and the ``dynamical''
$p^2$-dependence  tend to cancel each other partly at small $x<0.3$. At
$x\sim 0.5-0.6$ the effect of the $p^2$-dependence of the bound nucleon
structure function is clearly visible and leads to an enhancement of the
nuclear binding effect.

A result (for ${}^{40}$Ca) of the unified description which incorporates
both the shadowing effect at small $x$ and the binding, Fermi motion and
off-shell corrections at large $x$ is shown in Fig.~\ref{fig10}. While the
overall pattern of the data is quite well reproduced, we see that once the
off-shell $p^2$ dependence of the bound nucleon structure function is
included, there is some room for a possible small enhancement of $F_2^A$
due to nuclear pion cloud effects \cite{Kulagi89,EriTho83}  for $0.2 < x <
0.4$, which are omitted in the present calculations.

\section{Summary and conclusions}

We have presented a unified description of deep inelastic scattering on
nuclear targets which covers the whole region $10^{-3} < x\lesssim 1$ of
the Bjorken variable. Our starting point is a relativistic, covariant
formalism which makes use of the analytic properties of quark correlators.
In the infinite momentum frame we recover the usual parton model.  In the
laboratory system, which is the appropriate frame in which to investigate
nuclear structure functions, this approach naturally incorporates two basic
mechanisms, namely (I) scattering from quarks and antiquarks in the target
and (II) photon conversion into a quark-antiquark pair and subsequent
interactions of this pair with target constituents.

For small $x$, say below $x < 0.1$, the second one of these processes
dominates and produces shadowing.  At the momentum transfers, $Q^2$,
typical of the current experiments we find that only about half of the
observed shadowing comes from the coherent interaction of uncorrelated
$q\bar q$ pairs with target nucleons, the mechanism discussed in
ref.\cite{BrodLu90}.  The other half comes from the coherent scattering of
strongly correlated pairs, i.e.  vector mesons.

At larger values of the Bjorken variable ($x > 0.2$), scattering from the
quarks of a single bound nucleon dominates. The leading nuclear effects are
now binding and Fermi motion, as pointed out already in previous studies.
However, we find that the naive convolution formalism needs to be
generalized to incorporate off-shell effects characteristic of bound
nucleons. These off-shell effects are by no means small.  They reduce
$F_2^A(x)/A$ with respect to the free nucleon structure function $F_2^N(x)$
and effectively enhance the binding correction.

\bigskip
We thank W. Melnitchouk and A. W. Thomas for discussions and comments.

\begin{appendix}
\section*{Nonrelativistic Reduction of Matrix Elements}

Consider eq.(\ref{fA-fi}) written for the distribution function
${\cal F}_A(x)$. For the following it is convenient to multiply it by $x$.
We have:
\begin{equation}    \label{xfA-xfi}
x{\cal F}_A(x)=-i\int\! \frac{d^4p}{(2\pi)^4}
\sum_{i=0}^3 {\cal A}_i\,x'f_i(x',p^2) .
\end{equation}
where \vspace*{-\baselineskip}\vspace*{-\abovedisplayskip}
\begin{mathletters} \label{A-i}
\begin{eqnarray}
{\cal A}_0  &=& {p_+\over M}\,{\rm Tr}\,\bar G(p) \\
{\cal A}_1  &=& {p_+\over M^2}\,{\rm Tr}(\bar G(p) \overlay{\slash}p) \\
{\cal A}_2  &=& {\rm Tr}(\bar G(p)\gamma_+)\\
{\cal A}_3  &=& {p^{\alpha}q^{\beta}\over Mq_0}\,
{\rm Tr}\left(\bar G(p)\sigma_{\alpha\beta}\right) .
\end{eqnarray}
\end{mathletters}%
Eqs.(\ref{A-i}) are written in the laboratory frame where
$P_A=(M_A,{\bf 0})$, $q=(q_0,{\bf 0}_T,-|{\bf q}|)$, $p_+=p_0+p_3$ and
$\gamma_+=\gamma_0+\gamma_3$. We have used the fact that $x/x'=p_+/M$.
Also we have introduced $\bar G(p)=G(p,P_A)/2M_A$ where the factor
$2M_A$ ensures that $\bar G$ is independent of the normalization of
the nuclear state.

Consider a non-relativistic reduction of matrix elements (\ref{A-i}).
We start from the nucleon propagator which can be written as follows:
\begin{equation}    \label{Gbar}
\bar G(p)=-i\int dt\,
    e^{ip_0\cdot t}
    \langle T\left(N({\bf p},t)\bar N({\bf p},0)\right)\rangle ,
\end{equation}
where
$N({\bf p},t)=\int d^3{\bf r}\,\exp(-i{\bf p}\!\cdot\!{\bf r})\,N({\bf r},t)$
is the nucleon field operator in a mixed $({\bf p},t)$-representation and the
brackets denote the averaging over the nuclear ground state,
$\langle\cdots\rangle=\langle A|\cdots|A \rangle/\langle A|A \rangle$
(see also Section~\ref{frame}).

We apply a $1/c$-expansion technique to obtain an approximate solution of
the Dirac equation for the nucleon field $N$ in the non-relativistic limit
$|{\bf p}|/M\to 0$. Up to terms of order $1/c^3$ the nucleon field $N$ can
be written as follows:
\begin{equation}
\label{N-Psi}
N({\bf p},t)=C
\left(
\begin{array}{r}
\Psi({\bf p},t)\\
\frac{{\displaystyle\bf p}\mbox{\boldmath$\cdot\sigma$}}{\displaystyle 2M}
                \,\Psi({\bf p},t)
\end{array}
\right).
\end{equation}
We have introduced a two-component non-relativistic nucleon field $\Psi$.
The normalization constant $C$ is fixed by the charge (particle number)
conservation condition,
\begin{equation}
\int d^3{\bf r}\, N^{\dagger}({\bf r})N({\bf r}) =
\int d^3{\bf r}\, \Psi^{\dagger}({\bf r})\Psi({\bf r}) ,
\end{equation}
which gives $C=1-{\bf p}^2/8M^2$.

Now we are prepared to calculate traces (\ref{A-i}). We write the nucleon
four-momentum as $p=(M+\varepsilon, {\bf p})$. The squared four-momentum
is $p^2\simeq M^2+2M(\varepsilon-T)$ where $T={\bf p}^2/2M$ is the
nonrelativistic kinetic energy. We also introduce the nonrelativistic
nucleon  propagator
\begin{equation}    \label{G-nr}
{\cal G}(p)=-i\int dt\,
    e^{ip_0\cdot t}
    \langle T\left(\Psi({\bf p},t)\Psi^{\dagger}({\bf p},0)\right)\rangle ,
\end{equation}
Using eqs.(\ref{N-Psi},\ref{G-nr}) we have:
\begin{mathletters}
\label{trs}
\begin{eqnarray}
{\cal A}_0(p) &=& \sqrt{{p^2\over M^2}}\,(1+{p_3\over M})\,\mbox{tr}\,{\cal
G}(p), \\
{\cal A}_1(p) &=& {p^2\over M^2}\,(1+{p_3\over M})\,\mbox{tr}\,{\cal G}(p),\\
{\cal A}_2(p) &=& (1+{p_3\over M})\,\mbox{tr}\,{\cal G}(p) \\
{\cal A}_3(p) &=& i\frac{2\varepsilon\!-\!T}{2M^2}\,
\mbox{tr}\left[
{\cal G}(p)\left(\mbox{\boldmath$p$}\times\mbox{\boldmath$\sigma$}\right)_3
\right] ,
\end{eqnarray}
\end{mathletters}%
where the trace is taken with respect to spin. Corrections to
eqs.(\ref{trs}) are of order  $1/c^3$. From the last equation we see that
${\cal A}_3$ appears only in the order $|{\bf p}|^3/M^3$ and can therefore
be neglected at the present level of accuracy. Moreover this term vanishes
identically for spinless nuclei, where tr$\,{\cal
G}\mbox{\boldmath$\sigma$}=0$. Substituting eqs.(\ref{trs}) into
eq.(\ref{xfA-xfi}) and closing the integration contour in the upper half of
the complex $p_0$ plane we arrive at eq.(\ref{convol:nr}).

\end{appendix}
\newpage

\newpage
\begin{figure}
\caption{Two basic mechanisms of deep inelastic scattering.} \label{fig1}
\end{figure}

\begin{figure}
\caption{Compton amplitude to the leading order in $Q^2$.} \label{fig2}
\end{figure}

\begin{figure}
\caption{A schematic picture of the quark spectral function.} \label{fig3}
\end{figure}

\begin{figure}
\caption{The nucleon structure functions $F_2^N(x)$, $F_3^N(x)$ and the
antiquark distribution $\bar q(x)$. Experimental data for the $F_2^N(x)$
(circles) are from ref.\protect\cite{NMColl92}, for $F_3^N(x)$ (squares)
and $\bar q(x)$ (triangles) from ref.\protect\cite{Mattis90,CDHSW91}.}
\label{fig4} \end{figure}

\begin{figure}
\caption{The nucleon structure functions $F_2^N(x)$, $F_3^N(x)$ and $\bar
q(x)$ at large (a) and small (b) values of the Bjorken variable $x$.}
\label{fig5} \end{figure}

\begin{figure}
\caption{The $Q^2$-behavior of $F_2^N(x,Q^2)$ at small $x$.  The data
points are from ref.\protect\cite{NMColl92} at $x=0.008$ (circles) and
$x=0.0175$ (squares).  The contribution from vector mesons
(\protect\ref{VM}) is shown by the dashed line. The solid curve is the full
result using eq.(\protect\ref{F2Ntotal}). The dotted line is the result for
$F_2^N(x=0.01,Q^2)$  using an empirical parameterization of the NMC data
\protect\cite{NMColl92}.} \label{fig6}
\end{figure}

\begin{figure}
\caption{The shadowing effect calculated for ${}^{12}$C (a), ${}^{40}$Ca
(b) and ${}^{132}$Xe (c) in comparison with the experimental data
\protect\cite{NMColl91,E66592}.  The dashed curve represents a scaling part
of the shadowing. The solid curve is the result of the full calculation
including the vector mesons.} \label{fig7}
\end{figure}

\begin{figure}
\caption{Impulse approximation for the nuclear Compton amplitude.}
\label{fig8} \end{figure}

\begin{figure}
\caption{The ratio of the nuclear and nucleon structure functions
calculated using eq.(\protect\ref{expansion}) with
$\langle\varepsilon\rangle= -50\,$MeV and $\langle T\rangle=20\,$MeV.  The
dashed curve is the result without corrections due to the $p^2$-dependence
of the bound nucleon structure function.  The solid line is the result of
the full calculation. The experimental data for ${}^{197}$Au are taken from
\protect\cite{Arnold84}.}
\label{fig9} \end{figure}

\begin{figure}
\caption{Result of the full calculation for ${}^{40}$Ca, including
shadowing, binding and off-shell effects.}
\label{fig10} \end{figure}

\end{document}